\documentclass[%draft, %for not including graphics
BCOR2pt, %space hidden by book binding
captions=nooneline, %do not distinguish between one or more lines in captions
bibliography=totoc, %add bibliography to table of contents
numbers=noenddot, %removes points for special parts (e.g. appendix)
parskip=half, %leaves at least 1em of a row free before a new paragraph, also sets parindent to 0
headings=normal, %for smaller headings
abstracton %for abstract
]{scrartcl} %automatically a4paper, 11pt, oneside, titlepage

\usepackage[USenglish]{babel} %for US english line breaks
\usepackage{graphicx} %for including figures
\usepackage{tikz}
\usepackage{framed}
\usepackage{faktor}
\usepackage{floatrow}
\usepackage{remreset} %for resetting numberings, e.g. the footnote number
\usepackage[nouppercase]{scrlayer-scrpage} %package for page style with not only upper letters in the head
\usepackage{amsmath} %sophisticated mathematical formulas with amstex (includes \text{})
\usepackage{amssymb} %sophisticated mathematical symbols with amstex (includes \mathbb{})
\usepackage[thmmarks, 
amsmath, %for compatibility with amsmath
hyperref %for compatibility with hyperref
]{ntheorem} %for sophisticated theorem environments and proofs
\usepackage{bm} %for bold math symbols
\usepackage{bbm} %only for indicator functions
\usepackage{enumitem} %for automatic numbering of new enumerate environments
\usepackage[hang]{subfigure} %for subfigures
\usepackage{wrapfig} %for wrapping text around figures/tables
\usepackage{tabularx} %for special table environment (tabularx-table)
\usepackage{color}
\usepackage{algorithm}
\usepackage[noend]{algpseudocode}

\usepackage{dcolumn} %for aligning tabular entries with respect to a certain char, e.g. decimal point
\usepackage{booktabs} %for table layout
\usepackage{listings} %for including programming code
\usepackage{psfrag} %for editing text in (e)ps files
\usepackage[pdftex, %necessary for breakurl
setpagesize=false, %necessary in order to not change text-/paperformat for the document
pdfborder={0 0 0}, %removes border around links
pdfpagemode=UseOutlines, %top level bookmarks are visible
]{hyperref} %all links stay black and are thus invisible
\makeatletter
\newcommand\mytoday{\number\year-\ifcase\month\or 01\or 02\or 03\or 04\or 05\or 06\or 07\or 08\or 09\or 10\or 11\or 12\fi-\ifcase\day\or 01\or 02\or 03\or 04\or 05\or 06\or 07\or 08\or 09\or 10\or 11\or 12\or 13\or 14\or 15\or 16\or 17\or 18\or 19\or 20\or 21\or 22\or 23\or 24\or 25\or 26\or 27\or 28\or 29\or 30\or 31\fi} %see LaTeX book pg. 348
\makeatother
\setkomafont{sectioning}{\normalcolor\bfseries} %set headings as in standard class
\clearscrheadfoot %removes old settings for head and foot
\automark[section]{subsection} %automatically creates head text based on the subsections and, if no subsection there, sections
\setkomafont{captionlabel}{\bfseries} %uses \bfseries for the caption label
\newcolumntype{d}[2]{D{.}{.}{#1.#2}} %aligns the entry at "." of a column if "d" is used as column type where the arguments specify the number of digits to the left and right for which space is kept in a column (always set to the maximal numbers that appear in the whole table)
\newcommand*{\abstractnoindent}{} %define abstract such that it has no indent
\let\abstractnoindent\abstract
\renewcommand*{\abstract}{\let\quotation\quote\let\endquotation\endquote
\abstractnoindent}
\makeatletter
\renewcommand{\p@enumii}[1]{\theenumi(#1)}
\makeatother

\theoremstyle{break} %adds a newline after the heading of a theorem environment
\theoremheaderfont{\bfseries}
\theorembodyfont{}
\theoremseparator{}
\newtheorem{definition}{Definition}[section] %number all environments in a sequence (for every section) (order does not play a role)

\newtheorem{remark}[definition]{Remark}

\theoremstyle{nonumberbreak} %adds a newline after the heading of a theorem environment which is not numbered
\theoremsymbol{$\Box$}

\newcommand*{\IR}{\mathbb{R}}

\usetikzlibrary{snakes}
\usetikzlibrary{arrows,shapes,positioning}
\usetikzlibrary{decorations.markings}
\tikzstyle arrowstyle=[scale=2]
\tikzstyle directed=[postaction={decorate,decoration={markings,
    mark=at position 1 with {\arrow[arrowstyle]{stealth}}}}]

\begin{document}

% ändert die Figure-Überschriften
\renewcommand{\figurename}{Fig.}

\thispagestyle{plain}
	\begin{center}
		{\bfseries\Large Performance attribution with respect to interest rates, FX, carry, and residual market risks}
		\par
		\vspace{1cm}
		{\Large Jan-Frederik Mai\footnote{XAIA Investment GmbH, Sonnenstr.\ 19, 80331 M\"unchen, email: \texttt{jan-frederik.mai@xaia.com}}}
		\vspace{0.2cm}
		\\
	\end{center}
	\begin{center}
Version of \today.
\end{center}
\begin{abstract}
We develop a method to decompose the PnL of a portfolio of assets into four parts: 
\begin{itemize}
\item[(a)] PnL due to FX rate changes,
\item[(b)] PnL due to interest rate changes,
\item[(c)] carry gain due to time passing,
\item[(d)] PnL due to residual market risk changes (credit risk, liquidity risk, volatility risk etc.).   
\end{itemize}
We demonstrate the usefulness of our approach by decomposing the performance of an FX- and interest rate-hedged negative basis position in our fund XAIA Credit Basis II, and we apply the methodology to decompose the performance of our fund XAIA Credit Debt Capital in the first quarter of 2022 into PnL contributions of the single positions.
\end{abstract}

\section{Introduction}
Let $A_t$ denote the price of an asset at time $t$, and denote by $\chi_t$ the price of one unit of the asset's currency in EUR. We assume that the asset is subject to reinvestment, which means that potential coupon or dividend payments (if present at all) are immediately re-invested into the asset. But we comment on the general case of distributing assets, or even a whole portfolio of assets with frequent rebalancing, below in Section \ref{sec_general} as well. The value $A_t\,\chi_t$ equals the EUR price of the asset at time $t$. We are interested in a decomposition of the EUR PnL of the asset in a time period $(t,T]$. Intuitively, we analyze this PnL at the time point $T$, and the EUR PnL $P^{(A)}_{(t,T]}$ of the asset in the period $(t,T]$ is given by
\begin{gather}
P^{(A)}_{(t,T]} := A_T\,\chi_T-A_t\,\chi_t.
\label{PnL_def}
\end{gather} 
In order to measure the part of $P^{(A)}_{(t,T]}$ that is induced by changes in the FX rate $\chi$, we decompose $P^{(A)}_{(t,T]}$ as follows:
\begin{gather}
P^{(A)}_{(t,T]} = \underbrace{\frac{A_t+A_T}{2}\,(\chi_T-\chi_t)}_{=:{P}^{(A)}_{(t,T]}(\chi)}+\frac{\chi_t+\chi_T}{2}\,(A_T-A_t).
\label{decomp_FX}
\end{gather}
In the following paragraph, we further decompose $A_T-A_t$ into different parts. Section \ref{sec_reinvest} first treats the case of assets subject to reinvestment, while Section \ref{sec_general} treats the general case and shows how to apply the methodology for whole portfolios of assets. Finally, Section \ref{sec_examples} demonstrates our decomposition by an example and applies the methodology to a performance attribution for our fund XAIA Credit Debt Capital.

\begin{remark}[Alternative FX performance decompositions]\label{rmk_FX}
The decomposition of the PnL into FX-induced part and non-FX-induced part (\ref{decomp_FX}) intuitively assumes that (i) the whole FX rate difference $\chi_T-\chi_t$ is earned on the arithmetic average of the asset values $A_t$ and $A_T$ at the beginning and at the end of the period, and (ii) the performance due to asset value change $A_T-A_t$ is converted into EUR according to the arithmetic average of the initial and latest FX rates $\chi_t$ and $\chi_T$. An alternative could be to only use $A_t$ in (i) and $\chi_T$ in (ii), by replacing the decomposition (\ref{decomp_FX}) with the alternative definition
\begin{gather*}
P^{(A)}_{(t,T]} = \underbrace{A_t\,(\chi_T-\chi_t)}_{=:\tilde{P}^{(A)}_{(t,T]}(\chi)}+\chi_T\,(A_T-A_t).
\end{gather*}
While both definitions ${P}^{(A)}_{(t,T]}(\chi)$ and $\tilde{P}^{(A)}_{(t,T]}(\chi)$ appear to be reasonable, we prefer (\ref{decomp_FX}). We notice that this is different to the practice of using the column ``Result curr.\ global'' in the front office system SOPHIS for FX-hedged performance monitoring, because the latter essentially corresponds to the expression $\chi_T\,(A_T-A_t)$.
\par
Yet another, more theoretical, decomposition is obtained by assuming that both $\{A_u\}_{u \in (t,T]}$ and $\{\chi_u\}_{u \in (t,T]}$ are realizations of semi-martingales. The product formula then gives the identity
\begin{gather}
P^{(A)}_{(t,T]} = \int_{(t,T]}A_u\,\mathrm{d}\chi_u+\int_{(t,T]}\chi_u\,\mathrm{d}A_u + [A,\chi]_t.
\label{int_decomp}
\end{gather}
In intuitive terms, the right-hand side of (\ref{int_decomp}) is a limit expression as $n \rightarrow \infty$ of 
\begin{gather*}
\sum_{i=1}^{n}A_{u_{i-1}}\,(\chi_{u_i}-\chi_{u_{i-1}})+\sum_{i=1}^{n}\chi_{u_{i-1}}\,(A_{u_i}-A_{u_{i-1}})+\sum_{i=1}^{n}(A_{u_i}-A_{u_{i-1}})\,(\chi_{u_i}-\chi_{u_{i-1}}),
\end{gather*}
where the partition $t=u_0<u_1<\ldots<u_n=T$ of the interval $(t,T]$ becomes finer and finer with increasing $n$. Under the assumption that $A$ and $\chi$ are independent (or of finite-variation) the co-variation term $[A,\chi]_t$ vanishes and we observe that our decomposition (\ref{decomp_FX}) can be viewed as a rough proxy for the remaining two integrals, when the integrand in both terms is replaced by the arithmetic average of start and end value. In contrast, the integral expressions in formula (\ref{int_decomp}) take into account the whole path trajectories of $A$ and $\chi$ on the interval $(t,T]$, and may thus be considered closer to reality. If one is not willing to assume that $A$ and $\chi$ are independent, which is obviously an assumption that strongly depends on the considered asset $A$ in concern, an FX decomposition becomes more difficult. For instance, if we assume that there is a positive probability that $A$ and $\chi$ have common jumps and we observe such jump at $u \in (t,T]$, the expression $[A,\chi]_t$ contains $\Delta A_u\,\Delta \chi_u$. Which percentage of this expression should be attributed to the PnL induced by FX rate changes, and which to a change in the asset value? While this consideration based on the stochastic integral formula (\ref{int_decomp}) might appear quite academic and difficult to implement in practice, it makes very clear in what sense our definition (\ref{decomp_FX}) can only be an approximation to reality.
\end{remark}

\section{Single assets subject to reinvestment} \label{sec_reinvest}
We assume that the price $A_t$ is a function of three variables: (i) time $t$, (ii) a discounting curve $r_t$, (iii) additional market risk factors $x_t$. This means we have $A_t=A_t(r_t,x_t)$. The additional market risk factors $x_t$ depend on the specific asset and applied pricing model. For instance, in case of a bond or credit default swap $x_t$ might be a tuple consisting of a credit spread curve, a recovery rate assumption, and a liquidity spread (e.g.\ such as the negative basis as introduced in \cite{mai19}). The sub-index $t$ at the interest rate object $r=r_t$ and the market factor $x=x_t$ indicates that the states of these input variables depend on the time point. If we write $A_s(r_t,x_u)$ we price the asset at time $s$ with the discounting curve from time $t$ and the market factor variables at time $v$, noticing that this is a theoretical number that has not been observed at some time point (it is in general different from $A_s=A_s(r_s,x_s)$). Given this, we decompose the PnL due to a change in asset values as follows, taking into account the three different driving factors interest rate ($r$), market risk ($x$), as well as carry (due to time $t$ passing):
\begin{align*}
& A_T-A_t = A_T(r_T,x_T)-A_t(r_t,x_t) \\
& = \frac{A_T(r_T,x_T)-A_T(r_t,x_T)}{2}+ \frac{A_t(r_T,x_t)-A_t(r_t,x_t)}{2}&\mbox{(interest rate change)}\\
& +\frac{A_T(r_T,x_T)-A_T(r_T,x_t)}{2}+\frac{A_t(r_t,x_T)-A_t(r_t,x_t)}{2} &\mbox{(market risk change)}\\
& +\frac{A_T(r_T,x_t)-A_t(r_t,x_T)}{2}+\frac{A_T(r_t,x_T)-A_t(r_T,x_t)}{2}. &\mbox{(carry)}
\end{align*}
We obtain the following final decomposition of the EUR PnL into four parts:
\begin{align*}
& P^{(A)}_{(t,T]}  = \frac{A_t+A_T}{2}\,(\chi_T-\chi_t) & (=P^{(A)}_{(t,T]}(\chi))\\
& + \frac{\chi_t+\chi_T}{2}\,\Big(\frac{A_T(r_T,x_T)-A_T(r_t,x_T)}{2}+ \frac{A_t(r_T,x_t)-A_t(r_t,x_t)}{2}\Big) & (=:P^{(A)}_{(t,T]}(r))\\
& +\frac{\chi_t+\chi_T}{2}\,\Big(\frac{A_T(r_T,x_T)-A_T(r_T,x_t)}{2}+\frac{A_t(r_t,x_T)-A_t(r_t,x_t)}{2}\Big) &(=:P^{(A)}_{(t,T]} (x))\\
& + \frac{\chi_t+\chi_T}{2}\,\Big(\frac{A_T(r_T,x_t)-A_t(r_T,x_t)}{2}+\frac{A_T(r_t,x_T)-A_t(r_t,x_T)}{2} \Big). & (=:P^{(A)}_{(t,T]}(\mbox{carry}))
\end{align*}
The interpretation of $P^{(A)}_{(t,T]}(r)$ is the following: at an arbitrary time point $u$ the difference $A_u(r_T,x_u)-A_u(r_t,x_u)$ measures the PnL of the market price at that time that would be induced by a discounting curve change from $r_t$ to $r_T$. The PnL $P^{(A)}_{(t,T]}(r)$ is defined as the arithmetic mean of this difference for the two time points $u=t$ and $u=T$. The precisely same logic applies to the interpretation of $P^{(A)}_{(t,T]}(x)$, only with $r$ replaced by $x$. Finally, the PnL $P^{(A)}_{(t,T]}(\mbox{carry})$ intuitively should measure the change between the asset values $A_t$ and $A_T$ that is only due to time passing, without the effects of $r$ and $x$. Since the variables $r$ and $x$ change their values within the period $(t,T]$, one reasonable approach is to use an ``average'' of the variables $r,x$ on the period $(t,T]$. Since we only have $r,x$ available at the two time points $t$ and $T$, a pragmatic idea to accomplish such average is to mix the possible pairs $(r_u,x_s)$ for $u,s \in \{t,T\}$ in a way that is as ``neutral'' as possible. This is precisely what's done in the definition of $P^{(A)}_{(t,T]}(\mbox{carry})$.

\begin{remark}[On the approximative nature of our definitions]\label{rmk_r}
Similar as in Remark \ref{rmk_FX}, we point out that our definition of $P^{(A)}_{(t,T]}(r)$ and $P^{(A)}_{(t,T]}(x)$ in terms of an arithmetic average of start and end time point values is only a proxy to reality. Clearly, an average that would take into account all time points $u \in (t,T]$ would be more desirable from a theoretical perspective. For instance, based on the multivariate It\^{o} formula, under the assumption that $\{r_u\}_{u \in (t,T]}$ and $\{x_u\}_{u \in (t,T]}$ are realizations of semi-martingales, take values in $\IR$ (i.e.\ are not function-valued), and under the assumption that $r$ and $x$ are independent, we obtain the decomposition
\begin{align*}
A_T-A_t &= \Big(\int_{(t,T]}\frac{\partial}{\partial t}A_u(r_u,x_u)\,\mathrm{d}u\Big)\\
& \quad +\Big(\int_{(t,T]}\frac{\partial}{\partial r}A_s(r_u,x_u)\,\mathrm{d}r_u+\frac{1}{2}\,\int_{(t,T]}\frac{\partial^2}{\partial r^2}A_s(r_u,x_u)\,\mathrm{d}[r,r]_u\Big)\\
& \quad+\Big(\int_{(t,T]}\frac{\partial}{\partial x}A_s(r_u,x_u)\,\mathrm{d}x_u+\frac{1}{2}\,\int_{(t,T]}\frac{\partial^2}{\partial x^2}A_s(r_u,x_u)\,\mathrm{d}[x,x]_u\Big),
\end{align*}
and the three terms in $(.)$-brackets could be interpreted as performance due to carry, changes in $r$, and changes in $x$, respectively. While already this formula is difficult to implement in practice, we point out that typically $r_u$ is a function (an interest rate term structure), so that a generalization in this regard requires significantly more advanced stochastic integration techniques. Furthermore, the assumption of dependence between $r$ and $x$ induces additional quadratic co-variation terms to the last formula that must be dealt with, similar to the situation mentioned in Remark \ref{rmk_FX} when $A$ and $\chi$ have common jumps. While such more theoretical considerations lie beyond the scope of this more practically oriented article, we still find it important to highlight this issue, as it makes clear that our definition is only an approximation to reality. 
\end{remark}

\section{Distributing assets and whole portfolios} \label{sec_general}
In the logic of the aforementioned section it was important that the asset was subject to reinvestment, in the sense that it is not allowed that cash flows out of the asset, because this outflow would then be forgotten in the analysis. Now we take care about such potential outflows. Special attention is required in this case, because if at some time point $u$ between $t$ and $T$ a certain cash amount\footnote{Without loss of generality, only concerning terminology, we assume that $A$ is a credit instrument and $c_u$ is a coupon payment at coupon date $u$, but it could as well be that $A$ is a stock and $c_u$ a dividend payment.} $c_u$ flows out of the asset price $A_u$, so that the price process jumps down by $c_u$ at time $u$, then the amount $c_u$ must somehow find its way into the PnL. The simplest possibility is to simply add $c_u$ to the carry gain $P^{(A)}_{(t,T]}(\mbox{carry})$, and this is essentially what we do. Concretely, we assume a decomposition $t=u_0<u_1<\ldots<u_m=T$ of the interval $(t,T]$ with $u_0,\ldots,u_m$ being potential cash outflow dates (if no cash flows out at time $T$ we assume $c_m=0$). With $A_{u-}$ denoting the asset price an instant before some cash outflow time point $u$, we notice that $A_{u_i}=A_{u_i-}-c_{u_i}$, because the (dirty) asset price drops at time $u_i$ by the coupon amount $c_{u_i}$. 
With this terminology, the definition of $P^{(A)}_{(t,T]}(\mbox{carry})$ is enhanced to become
\begin{align}
P^{(A)}_{(t,T]}(\mbox{carry}) &= \sum_{i=1}^{m} \frac{\chi_{u_{i-1}}+\chi_{u_i}}{2}\,\Bigg(\frac{A_{u_i}(r_{u_i},x_{u_{i-1}})-A_{u_{i-1}-}(r_{u_i},x_{u_{i-1}})}{2} \nonumber\\
& \qquad +\frac{A_{u_i}(r_{u_{i-1}},x_{u_i})-A_{u_{i-1}-}(r_{u_{i-1}},x_{u_i})}{2} + c_{u_i}\Bigg),
\label{carryformula}
\end{align}  
and the total PnL is enhanced accordingly as well. 
\par
Here it is important to be aware that the FX rate $\chi$ enters at all coupon time points $u_i$, and this is a definition that one should ponder about for a little while. The given formula simply assumes that there is zero PnL between $u_i$ and $T$ arising from the coupon received at $u_i$, but the PnL on the received EUR amount is ``frozen'' at time $u_i$. Intuitively, at time $u_i$ the out-flowing EUR cash amount $c_{u_i}\,(\chi_{u_{i-1}}+\chi_{u_i})/2$ goes into some other asset and must be monitored in the PnL $P^{(\tilde{A})}_{(u_i,T]}$ for that other asset $\tilde{A}$. For instance, if nothing is done with this cash and it lays on some cash account during $(u_i,T]$, one must monitor the PnL on that cash account, but this is not done here in our formula, because it is dedicated to the asset under concern only. 
\begin{remark}[An alternative definition implemented in SOPHIS]
One potential alternative to the presented formula (\ref{carryformula}) could be to replace each occurrence of $\chi_{u_i}$ in the last term with $\chi_T$. Intuitively, this alternative has the appeal that $\chi_T$ can be factored out of the sum, and the remaining formula is totally FX-free within the asset's currency. Since this is what the front office system SOPHIS does in its column ``Result curr.\ global'', when the global currency is EUR, let us briefly comment on the implied change of meaning. When using $\chi_T$ always, the obtained PnL intuitively assumes that each received coupon amount $c_u$ at some time $u$ is held in cash (not on a bank account, and also not invested in some other asset) between $u$ and $T$. Changes in the FX rate at later time points then induce significant PnL movements to cash amounts that have long left the position under consideration. For example, consider a bond that pays a high coupon rate, so a lot of cash leaves the asset over time. But if the FX rate $\chi$ sharply increases at time $T$, all these past cash amounts induce PnL, even though they might have been consumed or spent elsewhere already and should not induce PnL at $T$ anymore.
\end{remark}
The change of the carry PnL part is especially important, if one aggregates several assets to a portfolio and wishes to decompose the PnL of the whole portfolio. It is then well possible that an asset is bought or sold at some time point $u$ in $(t,T]$. For instance, it could be that at time $u$ one receives a coupon payment from one asset and uses it to purchase another asset, or it could be that additional  money flows into (out of) the asset to increase (reduce) the holdings. In qualitative terms, this is similar to the aforementioned coupon cash outflows, because we have to deal with in- and outflows in our PnL measurement. In order to take such dynamic rebalancings of a whole portfolio $P$ into account in the PnL decomposition, we recommend to split the time period $(t,T]$ into several parts $t=:t_0<t_1<\ldots\leq t_n = T$, where the $t_i$ include \textbf{all} transaction dates of the portfolio, i.e.\ all dates at which an asset is either sold or bought. Then finally, we obtain the decomposition
\begin{align}
P^{(P)}_{(t,T]}&= \underbrace{\sum_{A \in P}\sum_{i=1}^{n-1}P^{(A)}_{(t_{i-1},t_i]}(\chi)}_{=P^{(P)}_{(t,T]}(\chi)}+\underbrace{\sum_{A \in P}\sum_{i=1}^{n-1}P^{(A)}_{(t_{i-1},t_i]}(r)}_{=P^{(P)}_{(t,T]}(r)}\\
& \qquad\qquad +\underbrace{\sum_{A \in P}\sum_{i=1}^{n-1}P^{(A)}_{(t_{i-1},t_i]}(x)}_{=P^{(P)}_{(t,T]}(x)}+\underbrace{\sum_{A \in P}\sum_{i=1}^{n-1}P^{(A)}_{(t_{i-1},t_i]}(\mbox{carry})}_{=P^{(P)}_{(t,T]}(\mbox{carry})} .
\end{align}
For the computation of this decomposition, required is availability of $\chi_t$, $A_t=A_t(r_t,\lambda_t,x_t)$, $r_t$, and $x_t$ at all time points $t_0,\ldots,t_n$, and for all assets $A$ in the portfolio $P$, as well as pricing routines that implement the asset prices in dependence on given parameters $r$, and $x$.
\section{Example and Application} \label{sec_examples}
\subsection{A negative basis example}
\begin{figure}[!ht]
 \begin{center}
    \includegraphics[width=\textwidth]{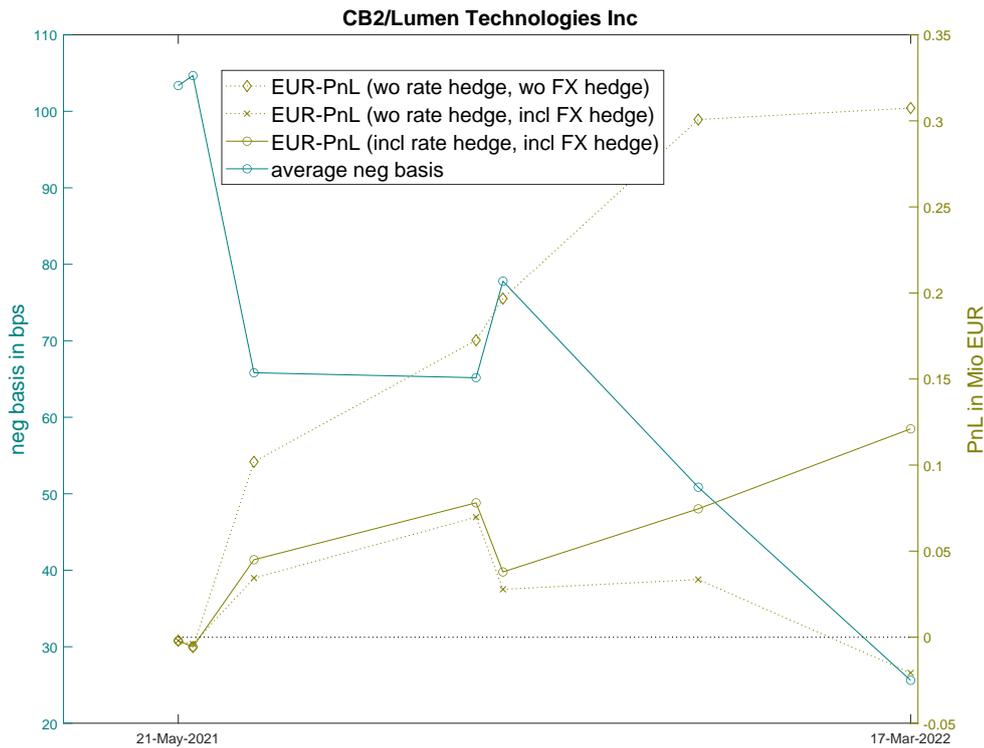}
    \end{center}
    \caption{The PnL of the negative basis position Lumen Technologies in our fund XAIA Credit Basis II from trade inception on 21 May 2021 until 17 March 2022. The PnL including both FX and interest rate hedge is essentially the position's contribution to the fund's performance. The PnL without hedge shows that the FX move was significant, and the PnL including FX hedge but without interest rate hedge is approximately equal to what we observe in our front office system SOPHIS in the column ``Result curr.\ global''. The depicted negative basis measurement is computed along the lines of \cite{mai19} and its decrease explains a huge part of the position's positive PnL.}
    \label{fig:Lumen}
\end{figure}

To demonstrate the methodology, we consider a negative basis position in our fund XAIA Credit Basis II on Lumen Technologies Inc. On 21 May 2021 we bought 4 mm USD bond and matched CDS protection (so-called bullet basis package). The EUR PnL of this position until 17 March 2022 is strongly positive by $313,710$ EUR, but mainly because the USD has appreciated with respect to the EUR, with the EUR-USD exchange rate falling from $1.22$ to $1.1$. Indeed, the number $P^{(A)}_{(t,T]}(\chi)$ equals $328,220$ EUR. The carry $P^{(A)}_{(t,T]}(\mbox{carry})$ equals $47,568$ EUR and the PnL due to market risk changes $P^{(A)}_{(t,T]}(x)$ amounts to $79,746$ EUR, so that their sum  $P^{(A)}_{(t,T]}(\mbox{carry})+P^{(A)}_{(t,T]}(x)$ equals around $127,310$ EUR. This PnL is essentially due to consumption and tightening of the negative bond-CDS basis, see \cite{mai19} for background. Since interest rates have risen significantly in the considered time period, the PnL $P^{(A)}_{(t,T]}(r)$ due to interest rate changes is negative and amounts to $-141,820$ EUR. Finally, the transaction costs due to the purchase on 21 May 2021 equals $6,268$ EUR in this example. In our fund XAIA Credit Basis II we hedge away interest rate risk and FX risk, in order to isolate the PnL $P^{(A)}_{(t,T]}(\mbox{carry})+P^{(A)}_{(t,T]}(x)$ (minus transaction costs). In the considered time period, our FX hedge has lost money and our interest rate hedge has gained money. The number $P^{(A)}_{(t,T]}(\mbox{carry})+P^{(A)}_{(t,T]}(x)$ (minus transaction costs) is around $121,042$ EUR, which is more than $3\%$ gain in nominal terms within less than one year. This is essentially the PnL from this trade that enters the fund's net asset value. In contrast, the (negative) PnL $P^{(A)}_{(t,T]}(\mbox{carry})+P^{(A)}_{(t,T]}(x)+P^{(A)}_{(t,T]}(r)$ (minus transaction costs) is essentially what's depicted in the folder ``Result curr.\ global'' in our front office system SOPHIS. Figure \ref{fig:Lumen} summarizes these numbers.

\subsection{An application to our fund XAIA Credit Debt Capital}
We present an application of the presented methodology to our fund XAIA Credit Debt Capital. In this fund, we seek to profit from relative valuation discrepancies between different assets referring to the same company. Concretely, the single investment positions can be categorized in one of the following buckets:
\begin{itemize}
\item \textbf{Capital Structure:} Long (or short) credit and short (or long) equity. A typical example is buying a bond (long credit) and buying an equity put option (short credit). Such positions are usually delta-neutral and gamma-long.
\item \textbf{Senior Sub:} Long senior credit and short subordinated credit. A typical example is buying a senior bond and buying subordinated CDS protection against it. Such positions are usually theta-negative and delta-short (meaning they profit from credit widening).
\item \textbf{Mismatch Basis:} Buy a long-dated bond and hedge it with shorter-dated CDS protection. Such positions are usually delta-neutral, gamma-positive and theta-negative, but can also be delta-long and theta-positive. Depending on the size of the bond-CDS-basis, these positions can sometimes even be theta-positive and gamma-long, which is our preferred case.
\item \textbf{Matched Basis:} Buy a bond and hedge it with maturity-matched CDS protection. Such positions are theta-positive and profit (suffer) from a tightening (widening) of the bond-CDS basis, see \cite{mai19} for background.
\item \textbf{Other:} Positions that cannot be attributed to some other bucket. Usually, these arise as left-overs of some closed positions, for instance a claim against a company after a bankruptcy.
\end{itemize} 
The fund eliminates interest rate- and FX- risk completely by hedges that are implemented on a global fund level. This means that the interest rate- and FX-deltas are aggregated over all positions, and the aggregated deltas are then neutralized by respective derivatives that are monitored in a separate bucket. On the one hand, this management practice is efficient because there is no need to enter hedging derivative contracts for each and every position. On the other hand, it is difficult to decompose the interest rate- and FX-hedges in the separate bucket in order to obtain attributions to the single positions. The latter difficulty is precisely what can be overcome with the presented methodology, which helps to isolate the actual contribution of the single positions to the total fund's PnL, just like in the negative basis example of the previous paragraph. 
\par
Table \ref{tab} depicts the resulting performance attribution for the first quarter of 2022 from 31 Dec 2021 until 01 Apr 2022. Within this quarter there have been significant movements in interest rates and in the EUR-USD exchange rate. For instance, the interest rate part $P^{(P)}_{(t,T]}(r)$ of all positions \textbf{without} consideration of the interest rate hedge bucket amounts to $-255$ bps of the fund's net asset value, due to sharp interest rate increases. Similarly, a significant appreciation of the USD with respect to the EUR in the considered period implied that $P^{(P)}_{(t,T]}(\chi)$ amounts to $120$ bps of the fund's net asset value. Consequently, the contributions of the two hedging buckets to the fund's performance was significant, and the attribution of this part of the PnL to the single positions is non-trivial. Especially positions with large bond exposures have had a poor performance due to rising interest rates, but these losses are compensated by the interest rate hedge bucket. Positions with a large USD exposure have had a strong performance due to an appreciation of the USD with respect to the EUR, but these gains correspond to respective losses in the FX hedge bucket. The presented methodology helps to split the significant contributions of the hedging bucket into parts associated to each single position. To this end, for each and every single position we simply compute the number $P^{(P)}_{(t,T]}(x)+P^{(P)}_{(t,T]}(\lambda)$ (minus transaction costs), which equals approximately the position's attribution to the total fund PnL \textbf{after} interest rate and FX hedge. 
\par
In Table \ref{tab}, the presented interest rate hedge costs result from both transaction costs and discretization as well as approximation errors, as mentioned in Remark \ref{rmk_r}. It appears to be significant in the considered period, which might be explained by the fact that there have been massive interest rate changes that are difficult to hedge accurately. It could also be partially explained by our definition of $P^{(P)}_{(t,T]}(r)$ based on averaging the beginning and the end of the periods. Possibly the actual loss due to interest rate changes was not as large as indicated by our methodology, which would then explain at least a part of the discrepancy. The FX costs/discrepancy includes FX hedging transaction costs, but may also partially be due to the fact that our definition of $P^{(P)}_{(t,T]}(\chi)$ does not match reality accurately, see Remark \ref{rmk_FX}. The given costs due to cash parking arise from non-invested cash that lies on overnight deposit accounts that cost around $78$ bps per annum in the considered time period. Fees include management fees ($50$ bps per annum) and performance fees ($20\%$ of performance exceeding 3M Euribor), and other costs comprise things like deposit charges and taxes. All presented numbers are approximate values.

\begin{table}[!htbp]
\begin{center}
\begin{tabular}{lllrrrr}
\specialrule{.1cm}{.2cm}{.2cm}
\multicolumn{3}{l}{\textbf{POSITIONS}}  &  &  &  & $\bm{127}$ \\
\specialrule{.1cm}{.2cm}{.1cm}
& \multicolumn{2}{l}{\textbf{Capital Structure}}   &  &  & $8$ &  \\
\specialrule{.05cm}{.1cm}{.1cm}
& & top: Microstrategy &  & {\scriptsize $29$}  &  &  \\
& & worst: Rite Aid Corp  & & {\scriptsize $-25$} &   &  \\
\specialrule{.05cm}{.1cm}{.1cm}
& \multicolumn{2}{l}{\textbf{Senior Sub}} & &   & $87$ &  \\
\specialrule{.05cm}{.1cm}{.1cm}
& & top: LOXAM   & & {\scriptsize $29$} & &  \\
& & worst: Picard   & & {\scriptsize $0$} &  &  \\
\specialrule{.05cm}{.1cm}{.1cm}
& \multicolumn{2}{l}{\textbf{Mismatch Basis}} &  &  & $30$ &  \\
\specialrule{.05cm}{.1cm}{.1cm}
& & top: Amkor Technology Inc & & {\scriptsize $7$} &  &  \\
& & worst: Limited Brands  & & {\scriptsize $-2$} &  &  \\
\specialrule{.05cm}{.1cm}{.1cm}
& \multicolumn{2}{l}{\textbf{Matched Basis}} &  &  & $2$ &  \\
\specialrule{.05cm}{.1cm}{.1cm}
& & top: Softbank  & & {\scriptsize $1$} &  &  \\
& & worst: KB Home  & & {\scriptsize $0$} &  &  \\
\specialrule{.05cm}{.1cm}{.1cm}
& \multicolumn{2}{l}{\textbf{Other}} & &  & $0$ &  \\
\specialrule{.05cm}{.1cm}{.1cm}
& & top=worst: Wirecard AG &  & {\scriptsize $0$} &  &  \\
\specialrule{.1cm}{.2cm}{.2cm}
\multicolumn{3}{l}{\textbf{CASH PARKING}}  &  &  &  & $\bm{-7}$ \\
\specialrule{.1cm}{.2cm}{.2cm}
\multicolumn{3}{l}{\textbf{FEES}}  & &  &  & $\bm{-30}$ \\
\specialrule{.1cm}{.2cm}{.2cm}
\multicolumn{3}{l}{\textbf{IR HEDGE COSTS}} &  &  &  & $\bm{-34}$ \\
\specialrule{.1cm}{.2cm}{.2cm}
\multicolumn{3}{l}{\textbf{FX COSTS/DISCREPANCY}} &  &  &  & $\bm{-4}$ \\
\specialrule{.1cm}{.2cm}{.2cm}
\multicolumn{3}{l}{\textbf{OTHER COSTS}} &  &  &  & $\bm{-3}$ \\
\specialrule{.2cm}{.2cm}{.2cm}
\multicolumn{3}{l}{\textbf{TOTAL}}  & &  &   & $\bm{49}$ \\
\specialrule{.2cm}{.2cm}{.2cm}

\end{tabular}
\end{center}
\caption{The presented performance attribution methodology is applied to decompose the PnL of our fund XAIA Credit Debt Capital within the first quarter of 2022, i.e.\ within the period from 31 December 2021 until 01 Apr 2022. The presented numbers are basis points relative to the fund net asset value, and the total fund performance within the considered period was 49 bps.}
\label{tab}
\end{table}%

\bibliographystyle{abbrv}

\end{document}